\documentclass[AMA,STIX1COL]{WileyNJD-v2}
\usepackage{moreverb}
\usepackage{bm}
\usepackage{setspace}
\usepackage{algorithm}
\usepackage{algpseudocode}
\usepackage{listings}
\usepackage{graphicx}
\usepackage[caption=false,font=footnotesize]{subfig}
\usepackage{verbatim}
\usepackage{hyperref}
\usepackage{makecell}
\usepackage{booktabs}
\usepackage{tikz,xcolor,hyperref}

\definecolor{lime}{HTML}{A6CE39}

\newcommand\BibTeX{{\rmfamily B\kern-.05em \textsc{i\kern-.025em b}\kern-.08em
T\kern-.1667em\lower.7ex\hbox{E}\kern-.125emX}}

\articletype{Research Article}%

\received{<day> <Month>, <year>}
\revised{<day> <Month>, <year>}
\accepted{<day> <Month>, <year>}

\begin{document}

\title{A Nearly Optimal Chattering Reduction Method of Sliding Mode Control With an Application to a Two-wheeled Mobile Robot}

\author[1]{Lei Guo*}

\author[1]{Han Zhao}

\author[1]{Yuan Song}

\authormark{LEI GUO \textsc{et al}}

\address[1]{\orgdiv{School of Artificial Intelligence}, \orgname{Beijing University of Posts and Telecommunications}, \orgaddress{\state{P.O. Box 108, No.10 Xi Tucheng Road, Haidian, Beijing}, \country{China}}}

\corres{*Lei Guo, Beijing University of Posts and Telecommunications. \email{guolei@bupt.edu.cn}}


\abstract[Abstract]{The problem we focus on in this paper is to find a nearly optimal sliding mode controller of continuous-time nonlinear multiple-input 
multiple-output (MIMO) systems that can both reduce chattering and minimize the cost function, which is a measure of the performance index of 
dynamics systems. First, the deficiency of chattering in traditional SMC and the quasi-SMC method are analyzed in this paper. In quasi-SMC, the signum function of 
the traditional SMC is replaced with a continuous saturation function. Then, a chattering reduction algorithm based on integral reinforcement 
learning (IRL) is proposed. Under an initial sliding mode controller, the proposed method can learn the nearly optimal saturation function using 
policy iteration. To satisfy the requirement of the learned saturation function, we treat the problem of training the saturation function as the 
constraint of an optimization problem. The online neural network implementation of the proposed algorithm is presented based on symmetric radius basis 
functions and a regularized batch least-squares (BLS) algorithm to train the control law in this paper. Finally, two examples are 
simulated to verify the effectiveness of the proposed method. The second example is an application to a real-world dynamics model---a 
two-wheeled variable structure robot.}



\maketitle

\section{Introduction}\label{sec:int}
%
%
%
%
\begin{figure*}[t]
  \centering
  \includegraphics[width=7in]{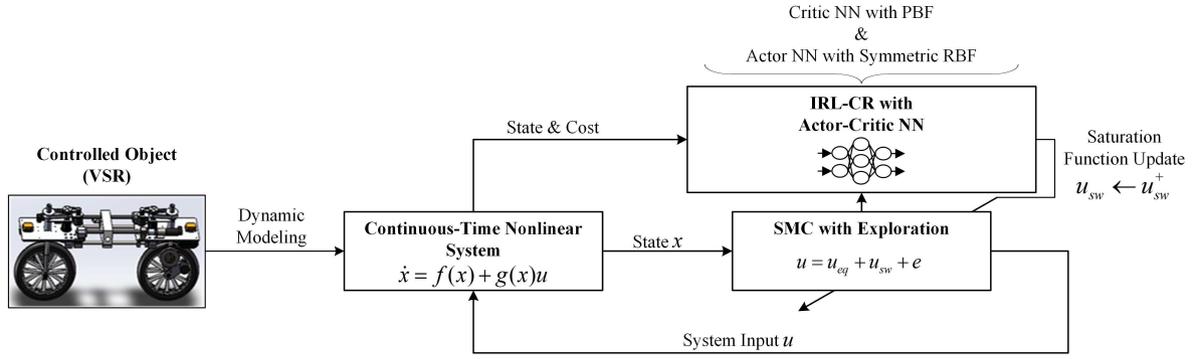}
  \caption{The control scheme of the method proposed in this paper.}
  \label{fig:scheme}
  \end{figure*}
    Sliding mode control (SMC), also known as variable structure control, is one of the most classic 
  feedback control algorithms. Unlike most control schemes, SMC has an infinite switching gain when the sliding 
  mode variable approaches zero. Therefore, the controller can be robust to small disturbances when the state variable 
  is close to the switching line (or manifold, more generally) $S=0$. However, the state variable will periodically 
  shuttle between two sides of the surface with high frequency and can never be strictly stabilized due to time-delays, 
  model uncertainties and other factors of the controllers and sensors, which is also known as chattering\cite{SMC}.  
    
    The chattering phenomenon could be very harmful in the real-world applications, as illustrated by the example of a two-wheeled mobile 
  robot designed in this paper. Similar to the inverted pendulum, the two-wheeled mobile robot has static instability 
  and dynamics stability, and the corresponding controller needs to adjust the posture of the robot in real time. Owing to the mechanical transmission 
  clearance, a high-frequency input signal cannot be perfectly executed by the actuator and can even damage it. Thus, the 
  chattering needs to be reduced when SMC is applied to the robot controller.

    The chattering reduction problem has been widely studied. The reaching control law was presented to reduce the amplitude 
  of chattering and maintain a fast approach speed to the switching manifold\cite{reaching}. A method that can 
  dynamically adjust the parameters of reaching law using fuzzy control was proposed in\cite{fuzzyreachinglaw}. In these methods, 
  the signum function still exists in the reaching law. Thus, the chattering cannot be completely eliminated.

    Several methods based on high-order terminal SMC have also been proposed\cite{terminalsmc1,terminalsmc2,terminalsmc3}. However, the 
  complexity of solving the control law and designing the switching manifolds limits their application to a real-world model. 
  The quasi-sliding mode method is the most commonly used controller for reducing chattering. The main concept of quasi-SMC is to 
  eliminate the discontinuous quantity of the signum function by replacing it with a continuous saturation function. A simple but effective 
  way is to choose the piecewise linear (PWL) function as the saturation function\cite{PWL} and to make the switching gain a linear 
  function of the sliding mode $S$ in a boundary layer. The method of changing the boundary layer is presented in\cite{adaptivePWL}. Similarly, 
  other forms of saturation functions have been studied, e.g. hyperbolic tangent functions \cite{tanhDC}. These functions are effective in 
  avoiding the chattering phenomenon but were designed considering only the stability analysis, with the parameters set by 
  experience and intuition. Little research has been focused on constructing and evaluating a standard performance index of quasi-sliding 
  mode controllers.

    The design of a saturation function such that the controller can simultaneously eliminate chattering and minimize the performance 
  index is worth studying. In the following sections, it is considered a constrained optimization problem. Traditional optimal 
  control methods, e.g., dynamic programming (DP)\cite{dp}, can theoretically solve the problem above. However, the DP algorithm must 
  solve the Hamilton-Jacobi-Bellman (HJB) equation (or Bellman equation in a discrete-time system), which makes it difficult or even impossible 
  to find the analytic solution when the model of a dynamics system is highly nonlinear, the state space has a high dimension, or the 
  dynamics model is not precisely known.

    Reinforcement learning (RL) is inspired by behavioral psychology and is a class of characteristic, data-driven algorithms 
  used to learn a nearly optimal policy by interacting with the environment\cite{introRL}. RL is also called adaptive dynamics 
  programming or approximate dynamic programming (ADP)\cite{adaptivedp} in the intelligent control field. In recent years, many RL 
  algorithms have been combined with deep learning by using deep neural networks as function approximators. Many deep RL methods have been proposed
  \cite{DQN,DDPG} and have efficient performances in different decision-making tasks\cite{alphago,anymal}. Unlike model-based 
  controllers such as SMCs and optimal controllers, model-free deep RL methods do not require any\emph{ a priori} knowledge of the system 
  dynamics. These algorithms, however, have not been proved to be convergent for continuous-time systems in the real-world\cite{IRLsurvey}. 

    To ensure that our controller of continuous-time (CT) systems is admissible and able to converge during the learning phase, 
  the controller design proposed in this paper is based on a class of ADP algorithms called integral reinforcement learning (IRL) algorithms\cite{IRL}, 
  which are used for solving the adaptive optimal control problem in CT dynamics systems, and its goal is to automatically learn an optimized 
  saturation function from online data, which can both improve the performance index and simulate some features of the signum function nicely.
    
    The rest of the paper is organized as follows. In Section \ref{sec:rel}, we discuss the mathematical form of sliding mode controllers 
  and a chattering reduction versus performance optimization dilemma. Finding a suitable saturation function is considered  
  an optimization problem in this section, and we introduce some features of the signum function as several constraint conditions. IRL algorithms 
  are also mentioned as related works in this section. The design of a chattering reduction controller based on an IRL algorithm, called integral 
  Q-learning I\cite{IRLexp}, is presented in Section \ref{sec:method}. The learning algorithm proposed in this section should be used with a stabilized initial 
  sliding mode controller to avoid massive inefficient exploratory actions and ensure convergence of the iterations. 
  An online implementation of the algorithm using neural networks is also formulated in this paper. By adding a penalty term to the objective 
  function and using a radius basis function neural network (RBFNN) to perform policy iteration, we transform the original problem 
  into an unconstrained one. Section \ref{sec:num} gives two examples to verify the effectiveness of this method. The second example 
  is a practical engineering application, that is, learning the controller of a variable structure wheeled robot working in 
  the inverse pendulum (Segway) mode. Finally, conclusions are drawn in Section \ref{sec:con}. The control scheme of the method proposed in this paper is shown in \figurename\ \ref{fig:scheme}.
    
    As for notations, we use $\Arrowvert{x}\Arrowvert$ to denote the Euclidean norm $\sqrt{x^Tx}$ of vector $x$. $X{\ }{\otimes}{\ }Y$ 
  indicates the Kronecker product of the matrices $X$ and $Y$ throughout this paper. The function of time $x(t)$ is denoted as $x_t$ or 
  $x$, and the function of other variables $f(x)$ will be written as $f$ in some equations for convenience.

  \section{Problem Formulation}\label{sec:rel}
  \subsection{Sliding Mode Control with Boundary Layers}\label{subsec:SMC}
    Consider the following continuous time system:
      \begin{equation}\label{MIMOsystem}
      \dot{x}(t)=f(x(t))+g(x(t))u(t),
      \end{equation}
  where $x{\ }{\in}{\ }\mathbb{R}^n$ is the state variable and $u{\ }{\in}{\ }\mathbb{R}^{m}$ is the control 
  input. It is assumed that system \eqref{MIMOsystem} can be stabilized on the set $\mathcal{D}{\ }{\subseteq}{\ }\mathbb{R}^n$ 
  containing the origin and that $f:\mathcal{D}{\ }{\to}{\ }\mathbb{R}^{n}$ with $f(0)=0$ and $g:\mathcal{D}{\ }{\to}{\ }\mathbb{R}^{n{\times}m}$ 
  are Lipschitz on $\mathcal{D}$.

    In conventional SMC, the switching manifolds are designed as straight lines. Thus, the sliding mode function is
      \begin{equation}\label{SMF}
      S=C^Tx=[s_1,s_2,...,s_m]^T,
      \end{equation}
  where $C{\ }{\in}{\ }\mathbb{R}^{n{\times}m}$ is a constant matrix. Each element denotes a straight line $s_i=0$
  in the state space:
      \begin{equation}\label{s_i}
        s_i=c_i^Tx.
      \end{equation}
  The sliding mode $S$ indicates the expected motion law of the state values. The state of the dynamics system reach the 
  switching manifold and move towards equilibrium.
  
    Under the ideal condition, i.e., the state values stay on the manifold $S=0$, the controller will ensure that the trajectories of 
  $x$ do not escape from the sliding surface. By solving the equation $\dot{S}=0$, we can obtain the equivalent control law $u_{eq}$, which is also the 
  general control law when $S=0$, if $(C^Tg)$ is a nonsingular matrix: 
      \begin{equation}
      \begin{split}
      \dot{S}&=C^T\dot{x}\\
             &=C^T(f(x)+g(x)u)=0\\
      &\Longrightarrow u_{eq}=-(C^Tg)^{-1}C^Tf.
      \end{split}
      \end{equation}

    A few forms of the reaching control law that can switch the state to the sliding mode $S=0$ has given in \cite{reachinglaw}. 
  Let $\dot{S}=-(WS+Ksgn(S))$, where $W=diag(w_i),K=diag(k_i)>0$ ($i=1,2,...,m$); the switching control law can be obtained 
  as
      \begin{equation}\label{uswpre}
      \begin{split}
      \dot{S}&=C^T(f(x)+g(x)u)=-(WS+Ksgn(S))\\
      &\Longrightarrow u_{sw}=-(C^Tg)^{-1}(WS+Ksgn(S)),
      \end{split}
      \end{equation}
  where $sgn(S)=[sgn(s_1),sgn(s_2),...,sgn(s_m)]^T$ is the signum function, denoted as
      \begin{equation}\label{sgn}
      sgn(s) = \left\{
      \begin{aligned}
        -1       {\ }{\ }{\ }{\ }{\ }{\ }& s<0\\
        0        {\ }{\ }{\ }{\ }{\ }{\ }& s=0\\
        1        {\ }{\ }{\ }{\ }{\ }{\ }& s>0
      \end{aligned}
      \right.,
      \end{equation}
  and \eqref{uswpre} is called the exponential velocity trending law.

    The general control law is the sum of the two control laws above:
      \begin{equation}\label{generalu}
      u=u_{eq}+u_{sw}=-(C^Tg)^{-1}(C^Tf+WS+Ksgn(S)).
      \end{equation}
  Take the Lyapunov function as $V=\frac{1}{2}S^TS$ to analyze the stability of the system. The derivative of $V$ with respect to time 
  is
      \begin{equation}\label{StabilityRequirement}
      \dot V=S^T\dot S=-WS^TS-K\sum_{i=1}^{m}\arrowvert{s_i}\arrowvert<0.
      \end{equation}
  Thus, the state variable can converge to the equilibrium point.

    The chattering phenomenon is mainly caused by the discontinuity of the signum function. An easy and effective way to solve 
  this problem is to replace $sgn(S)$ with a continuous saturation function:
      \begin{equation}
      sat(S)=[sat(s_1),sat(s_2),...,sat(s_m)]^T.
      \end{equation}

    The stability requirement \eqref{StabilityRequirement} can be satisfied if $sat(s_i)$ meets the 
  following conditions $(i=1,2,...,m)$:
    
    (a) $sat(s_i){\cdot}sgn(s_i)>0{\ }(s_i{\ }{\ne}{\ }0)$,

    (b) $sat(0)=0$.
    
    Three more conditions are proposed to simulate the signum function:
    
    (c) ${\lim_{s_i \to \infty}}sat(s_i)=sgn(s_i)$.

    (d) $sat(s_i)$ is an odd function.
    
    (e) ${\sup}_{s_i}|sat(s_i)|{\ }\le{\ }1$.

      \begin{figure}[t]
      \centering
      \includegraphics[width=3.4in]{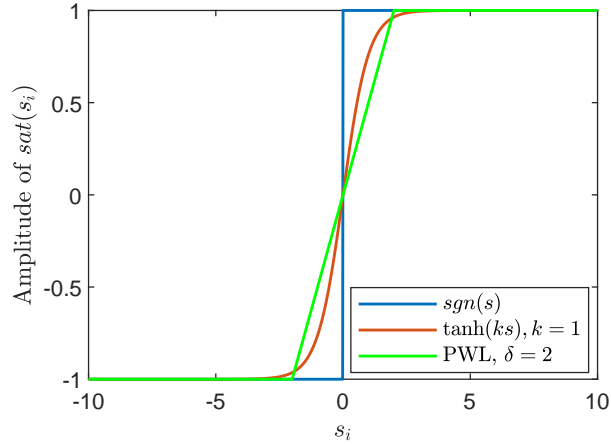}
      \caption{An example of different saturation functions: $sgn(s_i)$, $\tanh(ks_i)$ with $k=1$ and the PWL function with $\delta=2$.}
      \label{fig:presatfunc}
      \end{figure}

    The forms of the saturation function have been extensively studied. A piecewise linear (PWL) function is used 
in\cite{PWL} as the saturation function:
      \begin{equation}
      sat(s) = \left\{
      \begin{aligned}
      -1       {\ }{\ }{\ }{\ }{\ }{\ }& s<-{\delta}\\
      s/{\delta} {\ }{\ }{\ }{\ }{\ }{\ }& -{\delta}{\le}s{\le}{\delta}\\
      1        {\ }{\ }{\ }{\ }{\ }{\ }& s>{\delta}
      \end{aligned}
      \right.,
      \end{equation}
  where ${\delta}$ is a positive constant. In quasi-SMC with the PWL function as the saturation function, the domain $[-{\delta},{\delta}]$ is called the boundary layer. 
  It can be concluded that the reaching velocity ($\dot{S}$) is a linear function of $x$. Therefore, the state variable cannot reach the equilibrium point in 
  finite time, which means that the dynamics performance of this controller will be reduced as well. 
  
    The hyperbolic tangent function has also been chosen as a saturation function:
      \begin{equation}
        sat(s) = \tanh(ks),
      \end{equation}
  where $k>0$. An example of different saturation functions is shown in {\figurename} \ref{fig:presatfunc}.
  
    The effect of parameter $k$ was discussed in previous research by several simulations\cite{tanhDC}. However, the parameter was chosen by experience; 
  therefore, the performance index has not been optimized. It is necessary to achieve a balance between chattering reduction and dynamics performance 
  optimization.
    
  \subsection{Integral Reinforcement Learning}\label{subsec:IRL}
    In this subsection, we formulate the IRL methods based on the CT input-affine system \eqref{MIMOsystem}. IRL represents a class of 
  algorithms that find the nearly optimal controller for CT systems by learning from online data. Define the performance 
  index used in this paper as
    \begin{equation}\label{performanceindex}
    J(x_t,u) = \int_{t}^{\infty}r(x_\tau,u_\tau)d\tau,
    \end{equation}
  where $r(x,u)=Q(x)+u^TRu{\ }{\in}{\ }\mathbb{R}$ is the cost-to-go function with $Q(x){\ }{\ge}{\ }0$ and $R>0, {\forall}{\ }x{\ }{\in}{\ }\mathcal{D}$.
  The goal of both optimal control and the IRL algorithm is to find the optimal policy $\mu^*$ that can minimize the function $J$ along the trajectory of $x$. 

    The value function plays a central role in estimating the performance of a policy. In this paper, the 
  value function $V^{\mu}$ of an initial state with policy (control law) $u={\mu}(x)$ can be defined as
      \begin{equation}
      V^{\mu}(z):=J(z,{\mu}(z)),
      \end{equation}
  where $z{\ }{\in}{\ }\mathcal{D}$ is the initial state value $x(0)$ of the system. The domain $\Omega$ in state space $\mathcal{D}$ is called 
  the admissible region if for all $z{\ }{\in}{\ }\Omega$, the policy ${\mu}(x)$ is able to stabilize system \eqref{MIMOsystem}. 
  $\mu(x)$ is called an admissible policy in $\Omega$. With an admissible policy, $V^{\mu}(z)<{\infty}$ is satisfied, and $V^{\mu}({x})$ 
  can be considered a Lyapunov function of \eqref{MIMOsystem} in domain $\Omega$. From Bellman’s principle of optimality\cite{dp}, the optimal 
  policy $\mu^*(x)$ can be obtained as an equation denoted by the value function:  
      \begin{equation}
      \mu^*(x)=\arg\min_\mu V^{\mu}(x),
      \end{equation}
  where the value function of the optimal policy $V^{\mu^*}(x)$ is also the optimal value function that denotes the minimized performance 
  index:
      \begin{equation}
      V^{\mu^*}(x)=V^*(x)=\min_\mu V^{\mu}(x).
      \end{equation}

    Define the Hamiltonian as
      \begin{equation}\label{Hamiltonian}
      H(x,u,\lambda):=r(x,u)+p^T(f+gu).
      \end{equation}
  Take the multiplier as $p={\nabla}V$. The Hamiltonian can be regarded as a CT version of the Q-function\cite{Hamilton}, which is commonly used 
  to denote the value of different actions in discrete-time RL algorithms.

    An integral-temporal difference (I-TD) method to estimate the value function of policy ${\mu}({x})$ in a continuous-time model was proposed in \cite{IRL}. 
  The cost-to-go function can be written in interval form as:
      \begin{equation}\label{ITD}
      V^{\mu}({x(t+T)})-V^{\mu}({x(t)})=-\int_{t}^{t+T}r(x({\tau}),u({\tau}))d{\tau},
      \end{equation}
  where $[t,t+T]$ is a small-time interval and the value function can be obtained by solving the I-TD equation. 
  This estimation is also called a policy evaluation in RL algorithms.

    After the policy evaluation, the value function of the current policy is estimated. By solving the equation below, we find the maximum value of the Hamiltonian:
      \begin{equation}\label{partialH}
      \frac{\partial{H}}{\partial{u}}=2Ru+g^T{(\nabla}V^\mu)=0,
      \end{equation}
  and the greedy policy of this estimated value function is obtained as
      \begin{equation}\label{policyimprovement}
      \mu^{next}=-\frac{1}{2}R^{-1}g^T(x){\nabla}V^\mu(x).
      \end{equation}
  The solution to \eqref{policyimprovement} is called the policy improvement in the DP and RL algorithms. By repeatedly performing policy evaluation and improvement, we can cause the 
  value function and policy to converge to optimum if an initial admissible policy $\mu_0$ is applied for the first iteration\cite{IRL}. The iterative 
  updating formulas are obtained as
      \begin{equation}\label{I-PI}
      \left\{
      \begin{aligned}
      &V_{i+1}({x(t+T)})-V_{i}({x(t)})=-\int_{t}^{t+T}r(x({\tau}),\mu_i({\tau}))d{\tau}\\
      &\mu_{i+1}=-\frac{1}{2}R^{-1}g^T(x){\nabla}V_i
      \end{aligned}
      \right..
      \end{equation}
  After the value function converges to the optimum $V^*(x)$, the optimal policy is obtained:
      \begin{equation}\label{optimalpolicy}
      \mu^*(x)=-\frac{1}{2}R^{-1}g^T(x){\nabla}V^*(x).
      \end{equation}

    It is worth noting that the policy improvement step above requires\emph{ a priori} knowledge of the system drift dynamics $g(x)$. Integral Q-learning I, a 
  model-free method that can perform both policy evaluation and policy improvement without knowledge of the system dynamics $f(x)$ or $g(x)$, was proposed in\cite{IRLexp}. The exploration signal 
  $e$ is introduced to relax\emph{ a priori} knowledge requirement of the dynamics system:
    \begin{equation}\label{ex-system}
    \dot{x}=f(x)+g(x)(u+e),
    \end{equation}
  where $e$ is a nonzero exploration signal that satisfies the persistence of excitation (PE) condition\cite{PE}. Solve a single I-TD equation
    \begin{equation}\label{ITD-e}
    \begin{split}
    V_{i+1}({x(t+T)})&-V_{i+1}({x(t)})\\
    &=-\int_{t}^{t+T}\left[r(x,u)+2\mu^T_{i+1}Re_{\tau}\right]d{\tau},
    \end{split}
    \end{equation}
  the value function estimation and updating policy are obtained at the same time. The system will not be stable at the origin with exploration, which 
  enables the sensors to collect sufficient data online and perform policy improvement without explicitly using the information of $g(x)$. In other words, 
  integral Q-learning I is a completely model-free IRL method with\emph{ a priori} knowledge requirement of the initial admissible policy only\cite{IRLexp}. 

    The solution of \eqref{ITD-e} is approximated by collecting online data and minimizing the least-squares error. In these previous works, polynomial 
  regression neural networks based on batch least squares (BLS) were the most commonly used method to perform the approximation.
      
\section{IRL-Based Chattering Reduction Method}\label{sec:method}
  \subsection{Constrained Optimization Problem using IRL}\label{sec:constrained}
    In this section, we describe our work based on the two kinds of control algorithms mentioned above by combining model-based SMC and the model-free 
  approximation of value function and learned policy. Split the general control law \eqref{generalu} into the following two parts:
      \begin{equation}\label{up}
      \left\{
      \begin{aligned}
      &u_{p1}=-(C^Tg)^{-1}(C^Tf+WS)\\
      &u_{p2}=-(C^Tg)^{-1}Ksat(S)
      \end{aligned}
      \right..
      \end{equation}
  The signum function can be regarded as a discontinuous saturation function in $u_{p2}$.

    For system \eqref{ex-system}, the main goal of the chattering reduction method presented in this paper is to find the saturation function that optimizes the 
  performance index \eqref{performanceindex}. To satisfy requirements (a)--(e) mentioned in section \ref{sec:rel}, we set these conditions 
  as constraints of an optimization problem.
  
    The objective function with these constraints is denoted as
      \begin{equation}
      \label{optimizationproblem}
      \begin{split}
      &\mathop{\min}_{sat(S)}\int_{t}^{\infty}r(x_\tau,u_\tau)d\tau\\
      s.t.{\ }&sat(s_i){\cdot}sgn(s_i)>0{\ }(s_i{\ }{\ne}{\ }0),\\
              &{\lim_{s_i \to \infty}}sat(s_i)=sgn(s_i),\\
              &sat(s_i)=-sat(-s_i),\\
              &{\sup}_{s_i}|sat(s_i)|\le1,
      \end{split}
      \end{equation}
  where $u_\tau=u_{p1}+u_{p2}$.
    
    In this paper we try to find a solution to \eqref{optimizationproblem} using an algorithm based on integral Q-learning I. First, we consider the 
  unconstrained optimization problem described by $u_{p1}$ and $u_{p2}$. By substituting the general control law \eqref{generalu} into the equation above, 
  we can obtain the optimization problem with no constraints as the iterative equation below:
      \begin{equation}\label{ITD-it}
      \begin{split}
      V_{i+1}({x(t+T)})&-V_{i+1}({x(t)})+\int_{t}^{t+T}2u_{p2,i+1}^TRe_{\tau}d{\tau}\\
      &=-\int_{t}^{t+T}(Q(x)+u_{i}^TRu_{i}+2u_{p1}^TRe_{\tau})d{\tau},
      \end{split}
      \end{equation}
  where $u_{p2,0}=-(C^Tg)^{-1}Ksgn(S)$. During the iteration, the first part of the control law $u_{p1}$ is not updated. 

    The IRL-based chattering reduction algorithm is shown in Algorithm \ref{alg1}. However, this IRL method is applied 
  to solve the unconstrained optimal control problem. Thus, an IRL-based chattering reduction method is proposed in this paper 
  to meet the needs of problem \eqref{optimizationproblem}.
  
      \begin{algorithm}[t]
      \caption{IRL-Based Chattering Reduction}
      \label{alg1}
      \textbf{1. Initialization}:

      ${\ }{\ }{\ }$Find an initial control policy $u_0=u_{p1}+u_{p2,0}$ and let $i \gets 0$.

      \textbf{2. Online data collection}:
      
      ${\ }{\ }{\ }$Apply the sliding mode control policy $u_i$ and collect information for 
    ${\tau}=0{\to}NT$ with an initial state $x(0)=z$.

      \textbf{3. Policy evaluation \& improvement}:
      
      ${\ }{\ }{\ }$Solve \eqref{ITD-it}.

      \textbf{4. Stopping criterion}:
      
      ${\ }{\ }{\ }$Let $i \gets i+1$. Update policy $u_{p2,i}$ and initialize the state 
    $x{\ }{\gets}{\ }z$ and time $t{\ }{\gets}{\ }0$.
    
      ${\ }{\ }{\ }$Go to step 2, until
      \begin{equation}
      {\sup}_{x{\in}\mathcal{D}}\Arrowvert{sat_{i+1}(s_j)-sat_i(s_j)}\Arrowvert<{\varepsilon},j=1,2,...,m
      \end{equation}
      where ${\varepsilon}$ is a small, positive constant.
      \end{algorithm}

  \subsection{Online Implementation Using an Approximation NN}\label{subsec:onl}
    To find the solution of \eqref{ITD-it} under constraints, the value function and the policy, which is denoted by the saturation function, 
  can be parameterized as
      \begin{equation}
      \hat{V}(x)=\hat{\theta}_c^T{\psi}(x)
      \end{equation}
  and
      \begin{equation}
      \label{sats}
      sat(S)=\tanh(S)+{\Phi}^T(S)\hat{\theta}_a
      \end{equation}
  respectively, where 
  \begin{equation*}
    \tanh(S)=[\tanh(s_1),...,\tanh(s_m)]^T,
  \end{equation*}
  \begin{equation*}
    {\Phi}(S)=[{\Phi}^T(s_1),...,{\Phi}^T(s_m)]^T,
  \end{equation*}
  and
      \begin{equation*}
      \Phi(s_i) = [{\phi}_1(s_i),{\phi}_2(s_i),...,{\phi}_p(s_i)].
      \end{equation*}
  The value function $V(x)$ and the saturation function $sat(S)$ can be approximated by a critic NN and an actor NN, respectively. ${\psi}(x)$ 
  and ${\phi}_j(s_i)$ ($j=1,2,...,p$) are nonlinear basis functions. $\hat{\theta}_a$ and $\hat{\theta}_c$ are the 
  weight vectors. The hyperbolic tangent function $\tanh(s)$ used in\cite{tanhDC} can be regarded as a special case of function \eqref{sats} when 
  the dimension of the input vector $m=1$ and $\hat{\theta}_a$ is a null vector.

    In this paper, the quadratic polynomial basis function is used to approximate the value function, which can be obtained using 
  the Kronecker product:
      \begin{equation*}
      \begin{split}
      &\psi(x)=x{\ }{\otimes}{\ }x\\
      &=[x_1^2,x_1x_2,...,x_1x_n,x_2x_1,...,x_2x_n,...,x_n^2]^T.
      \end{split}
      \end{equation*}
    
    To make $sat(s_i)$ symmetric about the origin $s_i=0$ and satisfy the condition ${\lim_{s_i \to \infty}}
  sat(s_i)=sgn(s_i)$, the symmetry radial basis function (SRBF) is used in this paper:
      \begin{equation*}
      {\phi}_j(s_i)=e^{-{\gamma}_j(s_i-r_j)^2}-e^{-{\gamma}_j(s_i+r_j)^2},
      \end{equation*}
  where ${\gamma}_j$ and $r_j$ are both positive constants.

    \begin{figure}[t]
    \centering
    \includegraphics[width=3.4in]{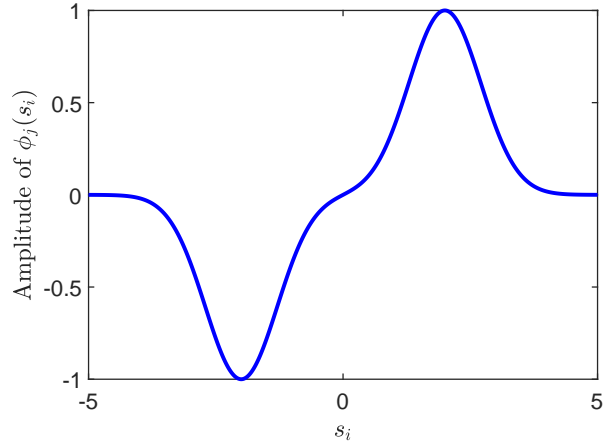}
    \caption{An example of the symmetric radial basis function where ${\gamma}_j=1$ and $r_j=2$.}
    \label{fig:SRBF}
    \end{figure}
    A single RBF neuron has the local response property: it can generate high-level gain only if the input is close to its center. 
  As shown in {\figurename} \ref{fig:SRBF}, the symmetry RBF will generate the maximum (or minimum) output if the input $s_i=r_j$ (or $-r_j$). 
  It is proved that the RBF network has the best ability to fit locally continuous nonlinear functions\cite{RBF}. Therefore, the RBFNN 
  may achieve satisfactory effects in approximating the saturation function within a boundary layer.

    The iterative equation is presented as:
      \begin{equation}\label{ITD-NN}
      \begin{split}
      &\hat{\theta}_{c,i+1}^T({\psi}(x(t+T))-{\psi}(x(t)))\\
      &+2\hat{\theta}_{a,i+1}^T\int_{t}^{t+T}{\Phi}(S)G^Te_{\tau}d{\tau}\\
      &=-\int_{t}^{t+T}2e_{\tau}^TG\tanh(S)+(Q(x)+u_{i}^TRu_{i}+2u_{p1}^TRe_{\tau})d{\tau}
      \end{split},
      \end{equation}
  where $G=R(C^Tg)^{-1}K$. After $N$ periods of data are collected, the $N$-dimensional linear equations can be obtained as
      \begin{equation}\label{BLSequation}
      {\Psi}_i\hat{\Theta}_{i+1}={\Xi}_i,
      \end{equation}
  where:
      \begin{equation*}
      \hat{\Theta}_{i+1}=[\hat{\theta}_{c,i+1},\hat{\theta}_{a,i+1}]^T,
      \end{equation*}
      \begin{equation*}
      {\Psi}_i=\left[\begin{matrix} {\psi}^T(x)|_t^{t+T}{\ }{\ }2\displaystyle{\int_{t}^{t+T}e_{\tau}^TG{\Phi}^T(S)d{\tau}}\\
                                    {\psi}^T(x)|_{t+T}^{t+2T}{\ }{\ }2\displaystyle{\int_{t+T}^{t+2T}e_{\tau}^TG{\Phi}^T(S)d{\tau}}\\
                                    {\vdots}{\ }{\ }{\ }{\ }{\ }\\
                                    {\psi}^T(x)|_{t+(N-1)T}^{t+NT}{\ }{\ }2\displaystyle{\int_{t+(N-1)T}^{t+NT}e_{\tau}^TG{\Phi}^T(S)d{\tau}}
                                    \end{matrix}\right],
      \end{equation*}
      \begin{equation*}
      {\Xi}_i=[{\xi}_1,{\xi}_2,...,{\xi}_N]^T
      \end{equation*}
        \begin{figure}[t]
        \centering
        \includegraphics[width=3.4in]{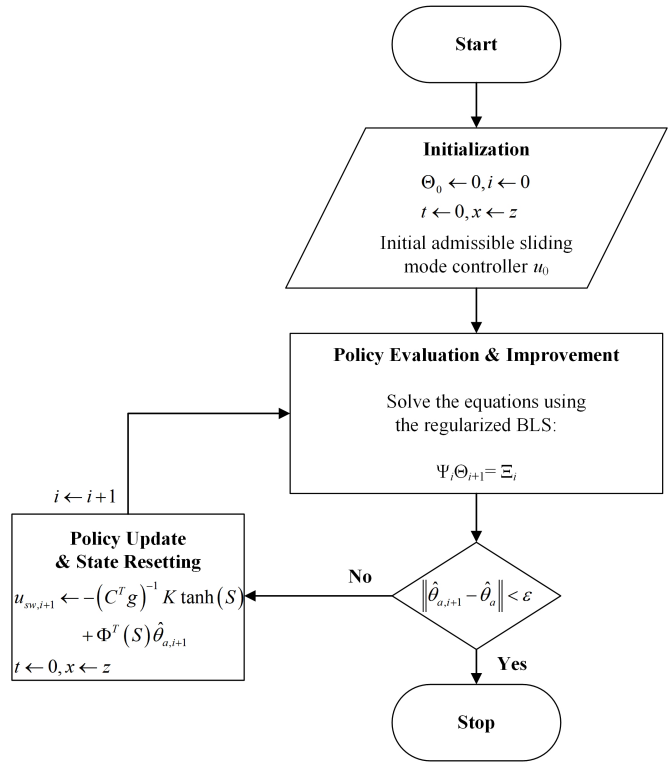}
        \caption{The flowchart of Algorithm \ref{alg1} using an approximation NN.}
        \label{fig:flowchart}
        \end{figure}
  and the elements of vector ${\Xi}_i$ can be obtained as:
      \begin{equation*}
      \begin{split}
      {\xi}_j&=-2\int_{t+(j-1)T}^{t+jT}e_{\tau}^TG\tanh(S)\\
             &-\int_{t+(j-1)T}^{t+jT}(Q(x)+u_{i}^TRu_{i}+2u_{p1}^TRe_{\tau})d{\tau},
      \end{split}
      \end{equation*}
  where $j=1,2,...,N$.
  
    Constraint conditions (a)--(d) in optimization problem \eqref{optimizationproblem} can be satisfied by using the RBFNN. The least-squares solution that 
  minimizes the approximation error can be obtained as follows using the BLS algorithm:
      \begin{equation}\label{BLSsolution}
      \hat{{\Theta}}_{i+1}=({\Psi}_i^T{\Psi}_i)^{-1}{\Psi}_i^T{\Xi}_i,
      \end{equation}
  where ${\Psi}_i^{\dag}=({\Psi}_i^T{\Psi}_i)^{-1}{\Psi}_i^T$ is the Moore-Penrose pseudoinverse matrix of ${\Psi}_i$. {\figurename} 
  \ref{fig:flowchart} shows the flowchart of our Algorithm \ref{alg1} using the approximation NN.

    The number of sample periods $N$ must be sufficient such that the following assumption is satisfied.
      \begin{assumption}\label{asu:N}
      The rank of matrix ${\Phi}_i$ meets the following condition:
      \begin{equation}\label{uniquesolution}
      rank({\Phi}_i){\ge}L_{\hat{\Theta}},
      \end{equation}
      where $L_{\hat{\Theta}}$ is the number of elements in vector $\hat{{\Theta}}$.
      \end{assumption}
    
    The ordinary BLS algorithm, however, cannot satisfy condition (e). In this paper, constraint (e) is modified as a penalty function 
  and the objective function can be rewritten as
    \begin{equation}
    \label{penalty}
    \begin{split}
    &\mathop{\min}_{sat(S)}\int_{0}^{\infty}r(x,u)d\tau+{\lambda}{\Arrowvert}\hat{\Theta}{\Arrowvert}^2\\
    s.t.{\ }&sat(S)=\tanh(S)+{\Phi}^T(S)\hat{\theta}_a,
    \end{split}
    \end{equation}
  where $\lambda>0$ is called the penalty term. Similar to \eqref{BLSsolution}, the weight updating formula using the penalty function 
  method is
    \begin{equation}\label{RBLSsolution}
    \hat{{\Theta}}_{i+1}=({\Psi}_i^T{\Psi}_i+{\lambda}I)^{-1}{\Psi}_i^T{\Xi}_i.
    \end{equation}
    
    The update rule above is called the $\ell_2$-regularization LS algorithm. The regularization method is also an effective way to improve the performance 
  of the approximator by avoiding overfitting in regression tasks\cite{Regularization}. The result of using the regularization method is shown 
  in the next section. The approximation can also be accomplished by the recursive least squares (RLS) algorithm, which can perform the policy evaluation step 
  at each time interval $T$. 
  
\section{Numerical Simulation}\label{sec:num}
    In this section, the chattering reduction method proposed in this paper is simulated with two examples to verify its effectiveness.
  \subsection{Example 1: Simple Single-input Single-output (SISO) System}\label{subsec:siso}
  \begin{figure}[t]
    \centering
    \includegraphics[width=3.4in]{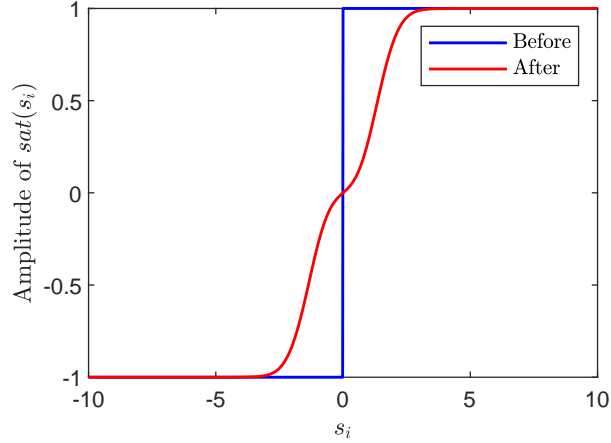}
    \caption{Example 1: Saturation function learned by the approximation NN using the regularization method. The weight vector in this 
  simulation is iterated 10 times.} 
    \label{fig:ex1sat}
  \end{figure}
  \begin{figure}[t]
    \centering
    \subfloat[]{\includegraphics[width=3.44in]{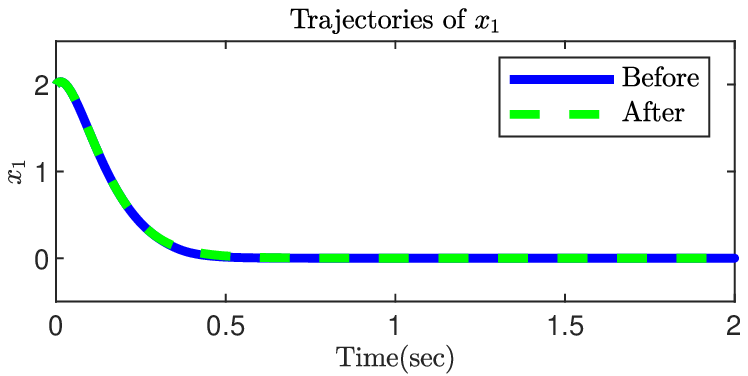} \label{subfig:x1ex1}}
    \subfloat[]{\includegraphics[width=3.44in]{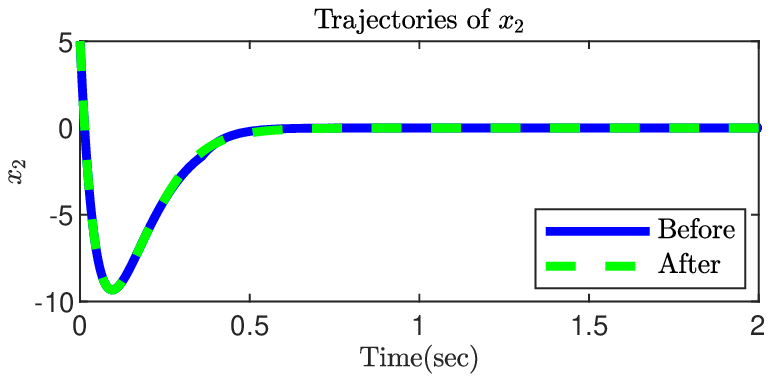} \label{subfig:x2ex1}}
    \quad
    \subfloat[]{\includegraphics[width=3.44in]{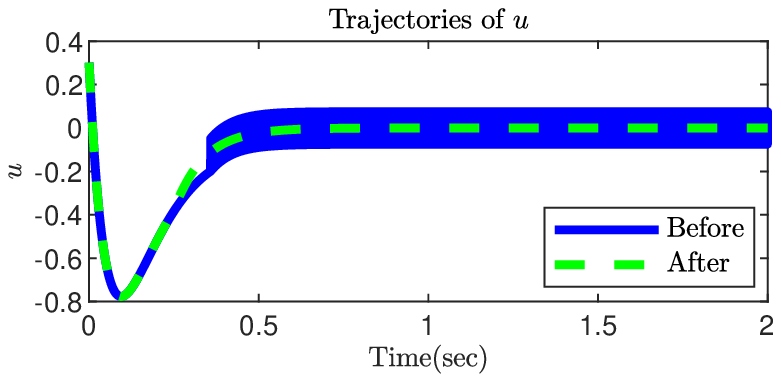} \label{subfig:uex1}}
    \caption{Example 1: Trajectories of the state variables (a) $x_1$, (b) $x_2$ and input (c) $u$. 
    Blue solid lines indicate the trajectories before using the chattering reduction algorithm, and the green dotted lines indicate the trajectories 
  after learning.}
    \label{fig:ex1tr}
  \end{figure}
    Consider an SISO linear system
      \begin{equation}\label{siso}
      \dot{x}=Ax+Bu,
      \end{equation}
  where 
      \begin{equation*}
        A=\left[\begin{matrix} 0&1 \\ 0&-25\end{matrix}\right], B=\left[\begin{matrix} 0 \\ 133\end{matrix}\right]
      \end{equation*}
  and the cost function is defined as 
      \begin{equation*}
        r(x,u)=10x_1^2+10x_2^2+u^2.
      \end{equation*}
  The sliding mode function can be obtained as $S=C^Tx$ where $C=[15,1]^T$. The other parameters in the sliding mode controller are as follows: $W=10$ and $K=10$, 
  and the initial control policy is
      \begin{equation}\label{ex1initial}
      \left\{
      \begin{aligned}
      &u_{p1}=-(C^TB)^{-1}(C^TAx+WS)\\
      &u_{p2,0}=-(C^TB)^{-1}Ksgn(S)
      \end{aligned}
      \right..
      \end{equation}

    In this example, the value function can be approximated by a linear combination of quadratic polynomials. We choose the basis of the critic NN as
      \begin{equation*}
      \begin{split}
      {\psi}(x)=[x_1^2,x_1x_2,x_2x_1,x_2^2]^T.
      \end{split}
      \end{equation*}
  
    The actor NN is initialized as a single-layer network with 7 symmetry RBF neurons. The hyperparameters ${\gamma}_j$ of the SRBFs 
  are all chosen as ${\gamma}_j=1$. The width of the boundary layer should be sufficiently small. Thus, the centers of these neurons we choose in this 
  example are near the origin and can be described as a vector
      \begin{equation*}
      \begin{split}
      vec(r_j)&=[r_1,r_2,...,r_7]^T\\
       &=[0.01,0.03,0.05,0.1,0.2,0.5,1]^T.
      \end{split}
      \end{equation*}
    
    Here, the number of sampling periods per iteration $N=200$, and the time interval $T=0.01$ s. The system starts at an initial 
  state $z=[2,5]^T$. The learning controller performs the policy evaluation and improvement every 2 seconds, and the states are reset to $z$ before the 
  next iteration. During the iteration, the exploration signal is set as $e=0.1\sin(\tau)$, and the regularization (penalty) term $\lambda=0.01$. 
        \begin{figure}[t]
          \centering
          \includegraphics[width=3.4in]{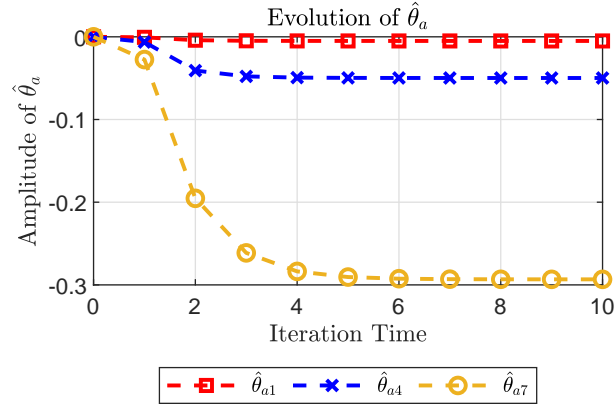}
          \caption{Example 1: Evolution of the weights $\hat{\theta}_{a1}$, $\hat{\theta}_{a4}$ and $\hat{\theta}_{a7}$ after 10 learning episodes.}
          \label{fig:ex1theta}
          \end{figure}
        \begin{figure}[t]
          \centering
          \includegraphics[width=3.4in]{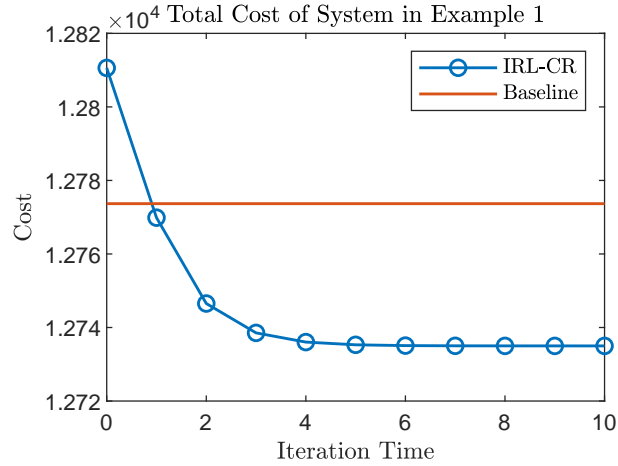}
          \caption{Example 1: The total cost of the system in each episode.}
          \label{fig:ex1cost}
          \end{figure}

    The saturation function learned by the NN after 10 iterations is shown in {\figurename} \ref{fig:ex1sat}. The weight vector 
  converges to 
      \begin{equation*}
      \begin{split}
      &\hat{\theta}_a=[-0.0050, -0.0150, -0.0250, -0.0498,\\
      &-0.0981, -0.2203, -0.2933]^T.
      \end{split}
      \end{equation*}

    The result shows the effectiveness of the presented algorithm in both decreasing the performance index along the trajectories and avoiding 
  chattering near the switching manifold. When an $\ell_2$ regularization term is added to \eqref{BLSsolution}, the regularized BLS
  method \eqref{RBLSsolution} minimizes the weighted sum of the approximation error and $\ell_2$-norm of the weight vector $\hat{\Theta}$. 
  {\figurename} \ref{fig:ex1cost} shows the effect of our method in decreasing the performance index. The blue line denotes the total cost 
  of our controller with exploration during the iteration. In this example, we choose the baseline to be the saturation function $sat(s_i)=\tanh(s_i)$, 
  which is presented in \cite{tanhDC}, to show the effectiveness of our method. This figure shows that the total cost of the system in each episode 
  monotonically decreases to convergence.
  
  \subsection{Example 2:Application to a Two-wheeled Variable Structure Robot (VSR)}\label{subsec:vsr}
  \begin{figure}[t]
    \centering
    \includegraphics[width=3.4in]{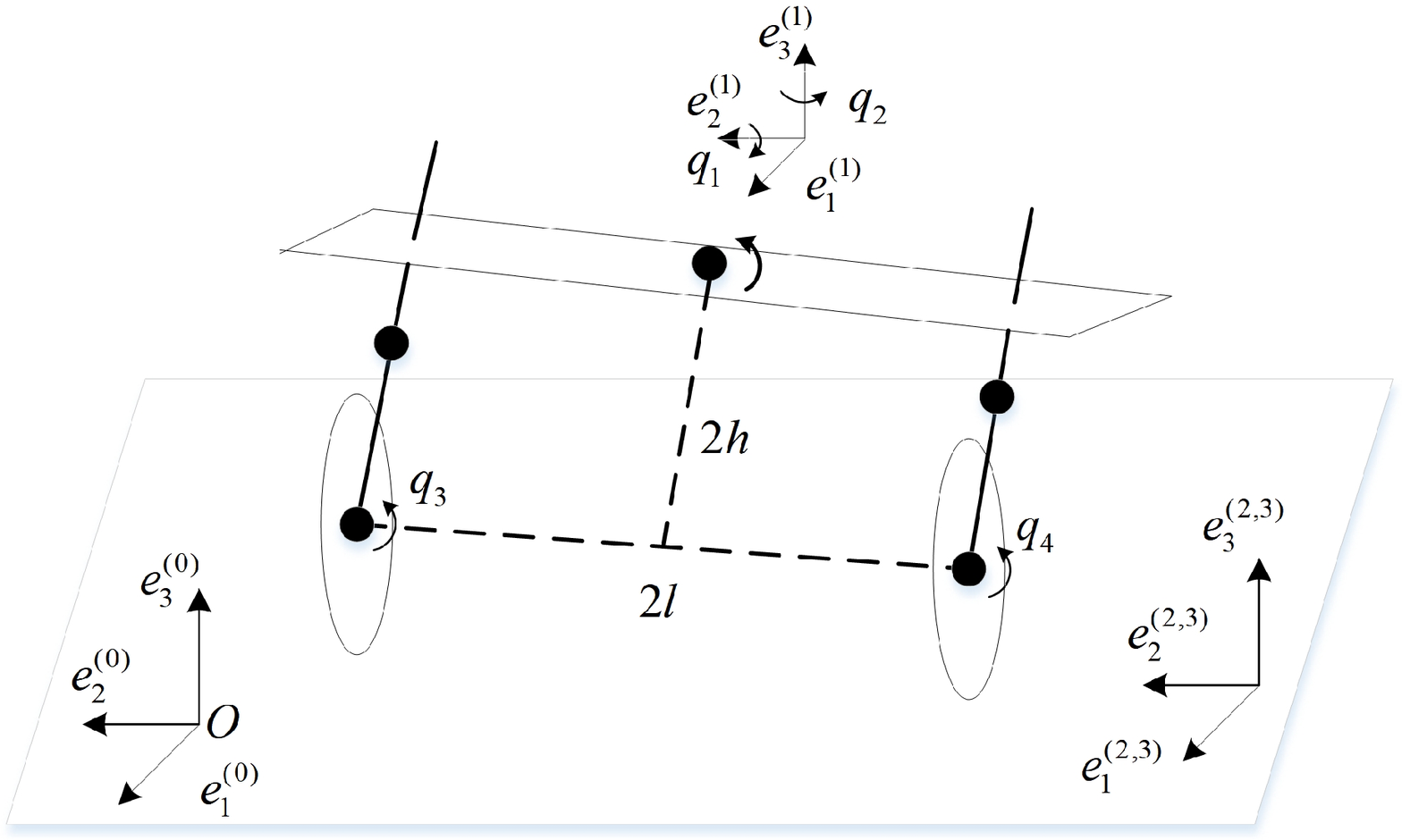}
    \caption{Example 2: Diagram of the variable structure robot.}
    \label{fig:vsr}
    \end{figure}
    \begin{figure}[t]
      \centering
      \includegraphics[width=3.4in]{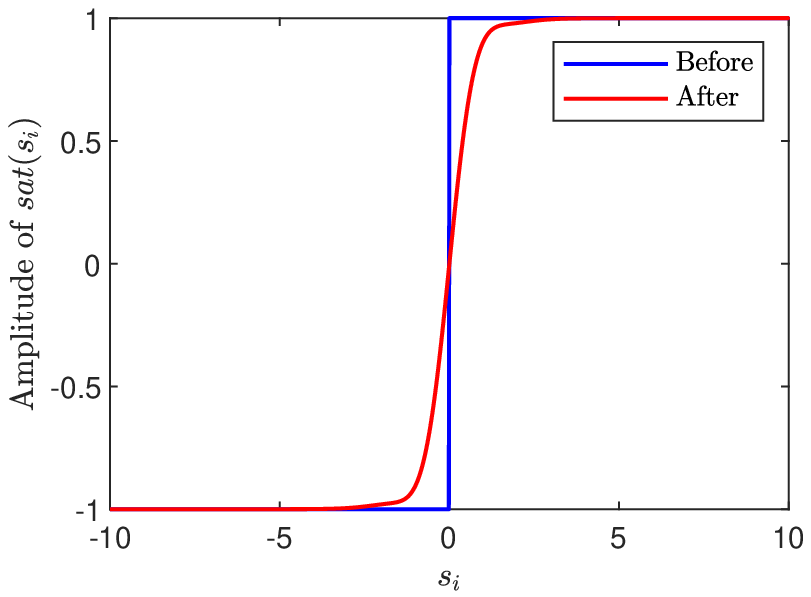}
      \caption{Example 2: Saturation function learned by the approximation NN using the regularization method. The weight vector in this 
    simulation is iterated 12 times.} 
      \label{fig:ex2sat}
    \end{figure}
    \begin{figure}[t]
      \centering
      \includegraphics[width=3.4in]{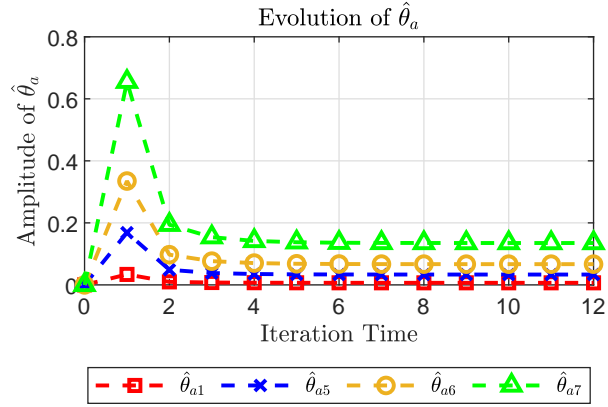}
      \caption{Example 2: Evolution of the weights $\hat{\theta}_{a1}$, $\hat{\theta}_{a5}$, $\hat{\theta}_{a6}$ and $\hat{\theta}_{a7}$ after 12 learning 
    episodes.}
      \label{fig:ex2theta}
    \end{figure}
    \begin{figure}[t]
      \centering
      \includegraphics[width=3.4in]{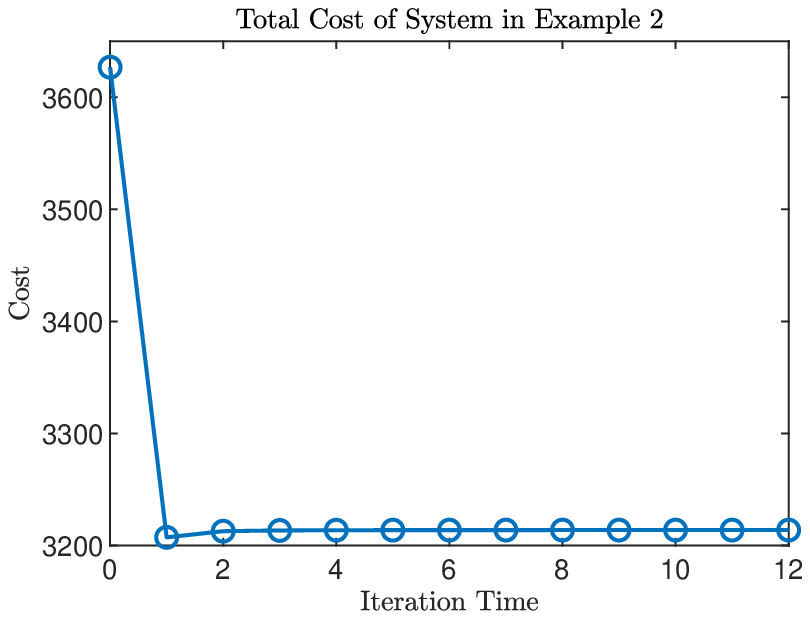}
      \caption{Example 2: The total cost of the system in each episode.}
      \label{fig:ex2cost}
      \end{figure}
    Consider an application to the dynamics model of our previous work, i.e., a two-wheeled VSR\cite{segway}. It is a wheeled robot that can 
  switch its structure between Segway mode, which can be regarded as an inverted pendulum, and self-driving bicycle mode. In this 
  paper we focus on the learning progress of a VSR working in Segway mode, which is designed for the robot to autonomously drive 
  under flat and broad road conditions.

  \begin{figure*}[t]
    \centering
    \subfloat[]{ \includegraphics[width=3.4in]{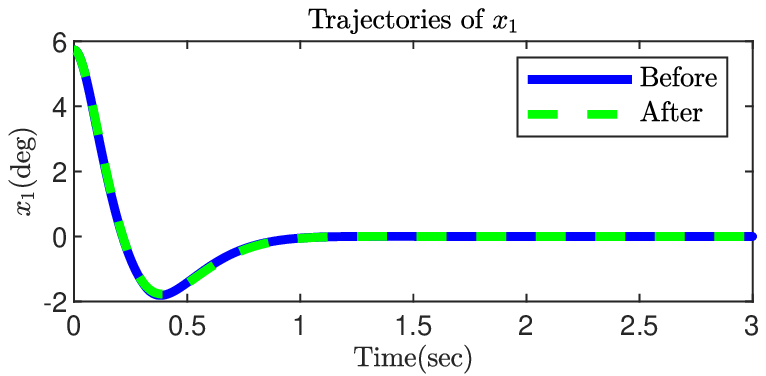} \label{subfig:x1ex2}}
    \subfloat[]{ \includegraphics[width=3.4in]{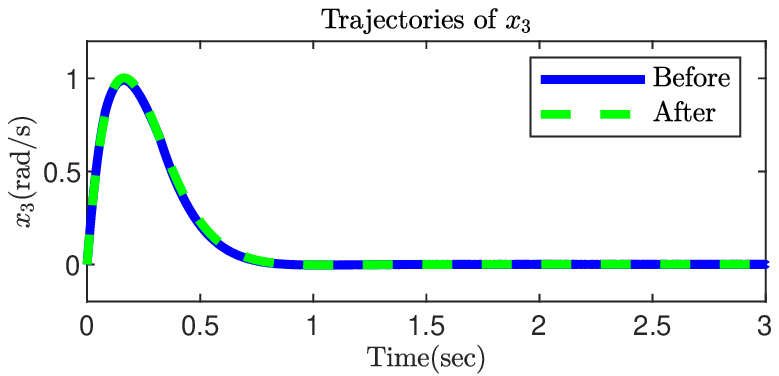} \label{subfig:x3ex2}}
    \quad
    \subfloat[]{ \includegraphics[width=3.4in]{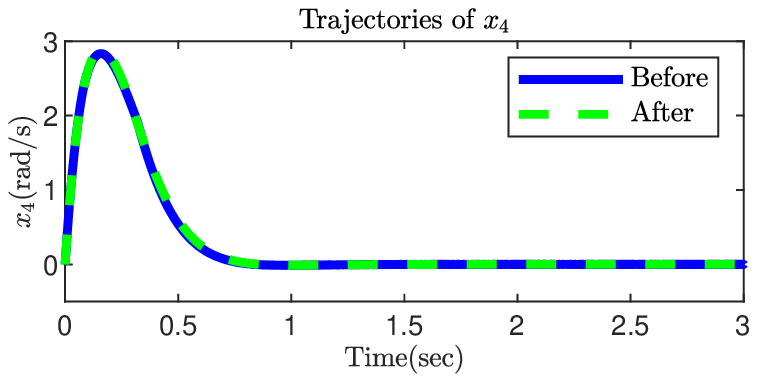} \label{subfig:x4ex2}}
    \subfloat[]{ \includegraphics[width=3.4in]{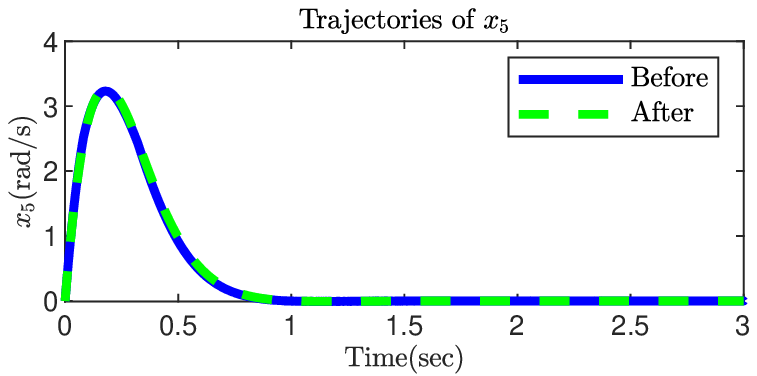} \label{subfig:x5ex2}}
    \quad
    \subfloat[]{ \includegraphics[width=3.4in]{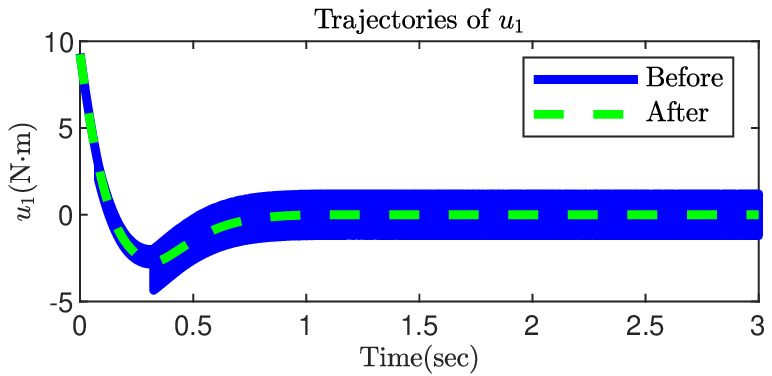} \label{subfig:u1ex2}}
    \subfloat[]{ \includegraphics[width=3.4in]{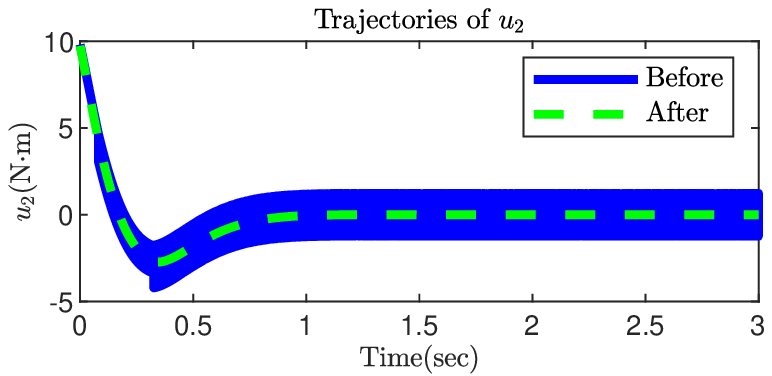} \label{subfig:u2ex2}}
    \caption{Example 2: Trajectories of the state variables (a) $x_1$, (b) $x_3$, (c) $x_4$, and (d) $x_5$ and the input (e) $u_1$ and (f) $u_2$.}
    \label{fig:ex2tr}
  \end{figure*}
    {\figurename} \ref{fig:vsr} shows the diagram of the VSR model. The task in this example is to stabilize the robot without considering the 
  motion path and the final position. The dynamics model of the VSR in Segway mode is obtained as
    \begin{equation}\label{VSRmodel}
      \left\{
      \begin{aligned}
      &\dot{x}_1=x_2\\
      &\dot{x}_2=145.1697{\sin}x_1-1.1790u_1-1.1790u_2\\
      &\dot{x}_3=-36.2317{\sin}x_1+0.7819u_1+0.7819u_2\\
      &\dot{x}_4=-55.5691{\sin}x_1+4.3571u_1-0.2575u_2\\
      &\dot{x}_5=-55.5691{\sin}x_1-0.2575u_1+4.3571u_2\\
      \end{aligned}
      \right.,
      \end{equation}
  where $x_1$ is the pitch angle of the robot and $x_2$ is the angular velocity of $x_1$. $x_3$ is the angular velocity of the yaw 
  angle. $x_4$ and $x_5$ are the angular velocity of the left- and right-wheel steering angles, respectively. To stabilize the state vector 
  $x=[x_1,x_2,x_3,x_4,x_5]^T$ to the equilibrium point 0, the coefficient matrix $C$ of the sliding mode function is chosen as
  \begin{equation*}
    C=\left[\begin{matrix} 80&3&2&-3&3 \\ 
                           1&6&2&3&-3\end{matrix}\right]^T
    \end{equation*}
  and the other parameters of SMC are designed as
      \begin{equation*}
      W=\left[\begin{matrix} 5&0 \\ 0&5\end{matrix}\right],K=\left[\begin{matrix} 50&0 \\ 0&50\end{matrix}\right].
      \end{equation*}

    The cost function is defined as 
    \begin{equation*}
      r(x,u)=\sum_{i=1}^5x_i^2+\sum_{j=1}^2u_j^2. 
    \end{equation*}
  In example 2, we built the actor NN with 8 different SRBF neurons, where the center vector was 
      \begin{equation*}
      vec(r_j)=[0.01,0.02,0.03,0.04,0.05,0.1,0.2,0.3]^T
      \end{equation*}
  and $\gamma_j=1$ ($j=1,2,...,8$). The number of sampling periods and the time interval were set as $N=300$ and $T=0.01$s, respectively. The penalty term $\lambda=0.001$. 
  As shown in {\figurename} \ref{fig:ex2theta}, after 12 iterations with the exploration signal set as
    \begin{equation*}
    e=[0.01\sin(\tau),0.01\sin(\tau)]^T,
    \end{equation*}
  the weight vector converged to 
      \begin{equation*}
      \begin{split}
      &\hat{\theta}_a=[0.0067,0.0133,0.0200,0.0267,\\
      &0.0333,0.0668,0.1350,0.2057]^T.
      \end{split}
      \end{equation*}
  The initial state was set as $z=[0.1,0,0,0,0]^T$ in this simulation. Similar to example 1, a data plot of the total cost is shown in {\figurename} \ref{fig:ex2cost}. The cost reached 
  its minimum after the first iteration. The weight vector $\hat\Theta$, however, was too large to satisfy condition (e) and was punished by the regularization term.
    
    To easily show the system response, we set the unit of $x_1$ in {\figurename} \ref{fig:ex2tr} \subref{subfig:x1ex2} as degrees instead of radians, 
  while the unit of the angular velocity was set as rad/s in the corresponding figure. As shown in {\figurename} \ref{fig:ex2tr}, the trajectories of the state variables nearly 
  identical in the simulations before and after learning. The chattering phenomenon of the input variables $u_1$ and $u_2$ disappeared after using the saturation 
  function learned from the sampling data.
  \begin{remark}
    In sections \ref{sec:rel} and \ref{sec:num}, we compared our method with different quasi-SMC methods. The saturation function learned by our algorithm consists of 
  the hyperbolic tangent function and several SRBFs. In theory, due to the differentiability and nonlinearity, our method can achieve better dynamics performance than 
  the PWL function\cite{PWL}. In section \ref{sec:num}, we also showed the advantage of our method by comparing it with the baseline\cite{tanhDC} in a simulation.
  \end{remark}
  \begin{remark}
    As mentioned above, the quasi-SMC using the hyperbolic tangent function\cite{tanhDC} is a special case of our method with the weight vector of the actor NN set as 0. Thus, 
  the iteration can also start with it instead of the traditional SMC. The simulation with two examples shows that our method can converge to the minimum value of the 
  performance index under the condition of satisfying the constraints we set in this paper. Due to the training period of NNs, the algorithm requires a sufficient computing 
  capability of the processer to ensure real-time computation online if the controlled object has a high-dimensional state and input space.
  \end{remark}
  \begin{remark}
    Our algorithm is based on the integral Q-learning I\cite{IRLexp}, which is an online policy iteration method. An initial admissible controller is needed in this method to 
  ensure the stability of the system. To satisfy the requirement, we set the initial controller as a model-based sliding mode controller. Thus, the algorithm proposed in this paper can be regarded 
  as a combination of pure model-based and model-free methods. The sliding mode controller designed offline can avoid numerous inefficient, meaningless exploration actions during the learning period, 
  while the model-free method can help the controller optimize the performance index.
  \end{remark}

  \section{Conclusions}\label{sec:con}
    The dilemma of reducing chattering and maintaining dynamics performance was discussed in this paper, and a chattering reduction method based 
  on IRL using an approximation NN was proposed. The major advantage of this new method is that it can reach a balance between improving the performance index and 
  reducing chattering. In section \ref{sec:num}, two simulation examples, especially the second one, verified the effectiveness of the proposed algorithm, and showed its 
  potential in real-world applications, e.g., as a stable, self-learning controller of robots. The combination of our work in this paper and high-order SMC 
  is a valuable research direction for the future. The robustness improvement of the proposed controller is also worth studying.

  \appendix{Dynamics Modeling of the VSR in Segway mode}
      \subsection{Notations \& Kinematic Analysis}
    In this paper, we define four different coordinate systems, which are also shown in {\figurename} \ref{fig:vsr}, to describe the motions of the robot:
  
    $e^{(0)}$ denotes the coordinate system fixed on the ground. Axis $e_1^{(0)}$ is vertical outward, $e_2^{(0)}$ is horizontal to the right 
  and $e_3^{(0)}$ is vertical up to the ground.

    $e^{(1)}$ denotes the coordinate system relative to the center of mass (CoM) of the bicycle frame.

    $e^{(2)}$ denotes the coordinate system relative to the CoM of the left wheel and the wheel rolls along the direction of $e_1^{(2)}$.

    $e^{(3)}$ denotes the coordinate system relative to the CoM of the right wheel and the wheel rolls along the direction of $e_1^{(3)}$.

    The physical parameters and notations of the states and inputs of the model are shown in {\tablename} \ref{tab:statesandinput} and \ref{tab:parameters}.

    The angular velocities of the bicycle frame, left fork and right fork are denoted as $\omega_1$, $\omega_2$, and $\omega_3$, respectively. In coordinate system 
  $e^{(0)}$, both of these parameters have the same value:
    \begin{equation}
      \omega_1=\omega_2=\omega_3=(0,-\dot{q}_1,-\dot{q}_2)^T.
    \end{equation}

    In Segway mode, the coordinate systems $e^{(1)}$, $e^{(2)}$, and $e^{(3)}$ are parallel to each other. The transformation matrices of coordinates 
  from $e^{(1)}$ to $e^{(2)}$ and $e^{(3)}$ are both three-dimensional unit matrix. Thus, the angular velocity of the left wheel in $e^{(2)}$ and the 
  angular velocity of the right wheel in $e^{(3)}$ are obtained as
    \begin{equation}
      \omega_4=\omega_2+(0,-\dot{q}_3,0)^T=(0,-\dot{q}_1-\dot{q}_3,-\dot{q}_2)^T,
    \end{equation}
  and
    \begin{equation}
      \omega_5=\omega_3+(0,-\dot{q}_4,0)^T=(0,-\dot{q}_1-\dot{q}_4,-\dot{q}_2)^T,
    \end{equation}
  respectively.

    It is assumed that both wheels perform a pure rolling motion on the ground. Thus, the velocity of the CoM of the left wheel in $e^{(2)}$ is
    \begin{equation}
      v_4=-\omega_4{\times}(0,0,r)^T=(r\dot{q}_1+r\dot{q}_3,0,0)^T.
    \end{equation}
  Similarly, the velocity of the CoM of the right wheel in $e^{(3)}$ is
    \begin{equation}
      v_5=-\omega_5{\times}(0,0,r)^T=(r\dot{q}_1+r\dot{q}_4,0,0)^T.
    \end{equation}

    The velocity of other components in coordinate system $e^{(1)}$ can be converted from $v_4$ and $v_5$. The velocity of the left fork and 
  the right fork can be obtained as
    \begin{equation}
      v_2=v_4+\omega_2{\times}(0,0,-h)^T=((h+r)\dot{q}_1+r\dot{q}_3,0,0)^T
    \end{equation}
  and
    \begin{equation}
      v_3=v_5+\omega_3{\times}(0,0,-h)^T=((h+r)\dot{q}_1+r\dot{q}_4,0,0)^T,
    \end{equation}
  respectively. The velocity of the frame is
    \begin{equation}
    \begin{split}
      v_1&=\frac{1}{2}(v_4+v_5)+\omega_1{\times}(0,-l,-2h)^T\\
         &=((2h+r)\dot{q}_1+l\dot{q}_2+\frac{1}{2}r(\dot{q}_3+\dot{q}_4),0,0)^T.
    \end{split}
    \end{equation}
      \subsection{Dynamics Analysis Using Chaplygin's Equation}
      Chaplygin's equation is a useful tool for performing a dynamics modeling, which is denoted as
      \begin{equation}
        \begin{split}\label{chaplygin}
        \frac{{{d}}}{{{{d}}t}}&\frac{{\partial \tilde T}}{{\partial {{\dot q}_\sigma }}} - \frac{{\partial \tilde T}}{{\partial {q_\sigma }}} - \frac{{\partial U}}{{\partial {q_\sigma }}}\\
        &+\sum\limits_{\beta  = 1}^\eta  {\frac{{\partial T}}{{\partial {{\dot q}_{\zeta  + \beta }}}}\sum\limits_{\gamma  = 1}^\zeta{\left({\frac{{\partial {B_{\zeta  + \beta ,\gamma }}}}{{\partial {q_\sigma }}} - \frac{{\partial {B_{\zeta+\beta,\sigma}}}}{{\partial{q_\gamma }}}}\right)}}{\dot q_\gamma}\\
        &={\tilde{Q}_\sigma}{\rm{(}}\sigma {\rm{ = 1,2,}}...{\rm{,}}\zeta {\rm{)}}
        \end{split},
      \end{equation}
    where $\zeta$ is the number of holonomic constraints, with $\zeta=4$ in Segway mode. $\eta$ denotes the number of nonholonomic constraints 
    of the system and $\eta=2$ in this model. $\tilde{T}$ or $T$ denotes the total kinetic energy with or without the nonholonomic constraints and 
    $U$ is the total potential energy of the system. $q_{\sigma}$ and $\tilde{Q}_{\sigma}$ are generalized independent coordinates and their corresponding 
    generalized moments of force. In nonholonomic constraint form $\dot{q}_{\zeta+\beta}$, the coefficient of $\dot{q}_{\gamma}$ is denoted as $B_{\zeta+\beta,\gamma}$.

    \begin{table}[!t]
    \renewcommand{\arraystretch}{1.3}
    \caption{State Variables and Inputs of the VSR}
    \label{tab:statesandinput}
    \footnotesize
    \centering
    \begin{tabular}{p{0.2\columnwidth}p{0.6\columnwidth}}
    \toprule
    \bfseries \makecell[c]{Notation}& \bfseries \makecell[c]{Meaning}\\
    \midrule
    \makecell[c]{$q_1$}&Pitch angle of the VSR.\\
    \makecell[c]{$q_2$}&Yaw angle of the VSR.\\
    \makecell[c]{$q_3$}&Roll angle of the left wheel.\\
    \makecell[c]{$q_4$}&Roll angle of the right wheel.\\
    \makecell[c]{$\tau_3$}&Torque the controller applies to the left wheel.\\
    \makecell[c]{$\tau_4$}&Torque the controller applies to the right wheel.\\
    \bottomrule
    \end{tabular}
  \end{table}
  
  \begin{table}[!t]
    \renewcommand{\arraystretch}{1.3}
    \caption{Physical Parameters of the VSR}
    \label{tab:parameters}
    \footnotesize 
    \centering
    \begin{tabular}{p{0.2\columnwidth}p{0.4\columnwidth}p{0.2\columnwidth}}
    \toprule
    \bfseries \makecell[c]{Notation}& \bfseries \makecell[c]{Meaning} & \bfseries \makecell[c]{Value}\\
    \midrule
    \makecell[c]{$J_{1x}$}&The moment of inertia (MoI) of the bicycle frame about the axis $e_1^{(1)}$.&\makecell[c]{1.18 kg$\cdot$m$^2$}\\
    \makecell[c]{$J_{1y}$}&The MoI of the bicycle frame about the axis $e_2^{(1)}$.&\makecell[c]{0.06 kg$\cdot$m$^2$}\\
    \makecell[c]{$J_{1z}$}&The MoI of the bicycle frame about the axis $e_3^{(1)}$.&\makecell[c]{1.21 kg$\cdot$m$^2$}\\
    \makecell[c]{$J_{2x}$}&The MoI of the bicycle fork about the axis $e_1^{(1)}$.&\makecell[c]{0.014 kg$\cdot$m$^2$}\\
    \makecell[c]{$J_{2y}$}&The MoI of the bicycle fork about the axis $e_2^{(1)}$.&\makecell[c]{0.017 kg$\cdot$m$^2$}\\
    \makecell[c]{$J_{2z}$}&The MoI of the bicycle fork about the axis $e_3^{(1)}$.&\makecell[c]{0.0046 kg$\cdot$m$^2$}\\
    \makecell[c]{$J_{3x}$}&The MoI of the wheel about the axis $e_1^{(2)}$ or $e_1^{(3)}$.&\makecell[c]{0.03 kg$\cdot$m$^2$}\\
    \makecell[c]{$J_{3y}$}&The MoI of the wheel about the axis $e_2^{(2)}$ or $e_2^{(3)}$.&\makecell[c]{0.05 kg$\cdot$m$^2$}\\
    \makecell[c]{$J_{3z}$}&The MoI of the wheel about the axis $e_3^{(2)}$ or $e_3^{(3)}$.&\makecell[c]{0.02 kg$\cdot$m$^2$}\\
    \makecell[c]{$r$}&The wheel radius.&\makecell[c]{0.17 m}\\
    \makecell[c]{$2h$}&The height difference between the CoM of a wheel and the CoM of the bicycle frame.&\makecell[c]{0.36 m}\\
    \makecell[c]{$2l$}&The distance between the CoMs of both wheels.&\makecell[c]{0.68 m}\\
    \makecell[c]{$m_1$}&The mass of the bicycle frame.&\makecell[c]{23.14 kg}\\
    \makecell[c]{$m_2$}&The mass of a bicycle fork.&\makecell[c]{0.5 kg}\\
    \makecell[c]{$m_3$}&The mass of a wheel.&\makecell[c]{3.8 kg}\\
    \bottomrule
    \end{tabular}
  \end{table}

      The nonholonomic constraint of the left wheel along $e_1^{(2)}$ and the right wheel along $e_1^{(3)}$ is obtained as
      \begin{equation}\label{cofleftw}
        \dot{q}_5=r\dot{q}_1+r\dot{q}_3
      \end{equation}
    and
      \begin{equation}\label{cofrightw}
        \dot{q}_6=r\dot{q}_1+r\dot{q}_4,
      \end{equation}
    respectively, due to the pure rolling motion assumption. The total kinetic energy, including the rotation and translation kinetic energy, is denoted as
      \begin{equation}\label{totalKEc}
        T=\sum_{i=1}^{5}(T_{1i}+T_{2i}){\ }{\ }{\ }(i=1,2,3,4,5),
      \end{equation}
    where the rotation kinetic energy is
      \begin{equation}\label{RKEc}
        T_{1i}=\frac{1}{2}J_i\omega^T\omega=\frac{1}{2}\left(J_{ix}\omega_{ix}^2+J_{iy}\omega_{iy}^2+J_{iz}\omega_{iz}^2\right).
      \end{equation}
    The translation kinetic energy of each component with the nonholonomic constraints is obtained as
      \begin{equation}\label{TKEc}
        \left\{
        \begin{aligned}
        &T_{21}=\frac{1}{2}m_1v_1^Tv_1\\
        &T_{22}=\frac{1}{2}m_2v_2^Tv_2\\
        &T_{23}=\frac{1}{2}m_2v_3^Tv_3\\
        &T_{24}=\frac{1}{2}m_3(\dot{q}_5^2+v_{4z}^2)\\
        &T_{25}=\frac{1}{2}m_3(\dot{q}_6^2+v_{5z}^2)
        \end{aligned}
        \right..
      \end{equation}
    The translation kinetic energy of the left and right wheels without nonholonomic constraints can be denoted as
      \begin{equation}\label{TKEl}
        \tilde{T}_{24}=\frac{1}{2}m_3v_4^Tv_4
      \end{equation}
    and
      \begin{equation}\label{TKEr}
        \tilde{T}_{25}=\frac{1}{2}m_3v_5^Tv_5,
      \end{equation}
    respectively. Substituting \eqref{totalKEc}--\eqref{TKEr} into \eqref{chaplygin}, the Chaplygin dynamics equations can be 
    obtained as \eqref{CE1}--\eqref{CE4}:
      \begin{equation}\label{CE1}
        \begin{split}
          0 &=  - 2gh({m_1} + {m_2})\sin {q_1}+ ({m_1} + 2{m_2} + 2{m_3}){r^2}){{\ddot q}_1}\\
          & + ({J_{1y}} + 2({J_{2y}} + {J_{3y}} + {h^2}(2{m_1} + {m_2})r + l{m_1}(2h + r){{\ddot q}_2}\\
          &+ \frac{1}{2}(2{J_{3y}} + 2h({m_1} + {m_2})r+({m_1} + 2{m_2} + 2{m_3}){r^2}){{\ddot q}_3} \\
          &+ \frac{1}{2}(2{J_{3y}} + 2h({m_1} + {m_2})r+({m_1} + 2{m_2} + 2{m_3}){r^2}){{\ddot q}_4}
        \end{split},
      \end{equation}
      \begin{equation}\label{CE2}
        \begin{split}
          0 &= l{m_1}(2h + r){\ddot q_1} + ({J_{1z}} + 2{J_{2z}} + 2{J_{3z}} + {l^2}{m_1}){\ddot q_2}\\
          & + \frac{1}{2}l{m_1}r{\ddot q_3} + \frac{1}{2}l{m_1}r{\ddot q_4}
        \end{split},
      \end{equation}
      \begin{equation}\label{CE3}
        \begin{split}
          {\tau _3} &= \frac{1}{2}(2{J_{3y}} + r(2h({m_1} + {m_2}) + ({m_1} + 2{m_2} + 2{m_3})r)){{\ddot q}_1} \\
          &+ \frac{1}{2}l{m_1}r{{\ddot q}_2} + ({J_{3y}} + (\frac{1}{4}{m_1} + ({m_2} + {m_3})){r^2}){{\ddot q}_3} + \frac{1}{4}{m_1}{r^2}{{\ddot q}_4}
        \end{split},
      \end{equation}
      \begin{equation}\label{CE4}
        \begin{split}
          {\tau _4} &= \frac{1}{2}(2{J_{3y}} + r(2h({m_1} + {m_2}) + ({m_1} + 2{m_2} + 2{m_3})r)){{\ddot q}_1}\\
          & + \frac{1}{2}l{m_1}r{{\ddot q}_2} + ({J_{3y}} + (\frac{1}{4}{m_1} + ({m_2} + {m_3})){r^2}){{\ddot q}_4} + \frac{1}{4}{m_1}{r^2}{{\ddot q}_3}
        \end{split}.
      \end{equation}
    
    By substituting the values of the parameters and performing identical deformations, we can obtain the dynamics model of the VSR \eqref{VSRmodel}.

\bibliography{bibfile}

\end{document}